\documentstyle[12pt]{article}

\newcommand{\nc}{\newcommand}
\def\frac#1#2{{\textstyle {#1 \over #2}}}

\nc{\beq}{\begin{equation}}
\nc{\eeq}{\end{equation}}
\nc{\beqa}{\begin{eqnarray}}
\nc{\eeqa}{\end{eqnarray}}
\nc{\lsim}{\begin{array}{c}\,\sim\vspace{-21pt}\\< \end{array}}
\nc{\gsim}{\begin{array}{c}\sim\vspace{-21pt}\\> \end{array}}

\def\&{and}

\def\nc#1#2#3{           {\it Nuovo Cim.  }{\bf #1}, #2 (19#3)}
\def\np#1#2#3{           {\it Nucl. Phys. }{\bf #1}, #2 (19#3)}
\def\pl#1#2#3{           {\it Phys. Lett. }{\bf #1}, #2 (19#3)}
\def\pr#1#2#3{           {\it Phys. Rev. }{\bf #1}, #2 (19#3)}
\def\prep#1#2#3{         {\it Phys. Rep. }{\bf #1}, #2 (19#3)}
\def\prl#1#2#3{          {\it Phys. Rev. Lett. }{\bf #1}, #2 (19#3)}

\begin{document}

\begin{titlepage}

\begin{center}
\hfill       YCTP-P13-96\\
\vskip .5 in
{\Large \bf  Phase Transitions in  Softly Broken
N=2 SQCD
at Non-Zero $\theta$ Angle }
\vskip .3 in

{{ Nick Evans\footnote{nick@zen.physics.yale.edu},
Stephen D.H. Hsu\footnote{hsu@hsunext.physics.yale.edu}
    and Myckola Schwetz\footnote{ms@genesis2.physics.yale.edu}
}}

   \vskip 0.3 cm
   {\it Sloane Physics Laboratory,
        Yale University,
        New Haven, CT 06511}\\
\end{center}

\vskip .5 in
\begin{abstract}
\noindent
We investigate the behavior of softly broken $N=2$ SQCD
at non-zero bare theta angle $\theta_0$, using
superfield spurions
to implement the SUSY breaking.
We find a first-order phase
transition as $\theta_0 $ is varied from
zero to $2 \pi$, in agreement with a
prediction of `t Hooft.
The  low-energy theta angle $\theta_{eff}$,
which determines
the effective charges of dyonic excitations, has
a complicated dependence on $\theta_0$ and
breaking parameters.

\end{abstract}
\end{titlepage}

\input epsf
\newwrite\ffile\global\newcount\figno \global\figno=1
\def\writedefs{\immediate\openout\lfile=labeldefs.tmp \def\writedef##1{%
\immediate\write\lfile{\string\def\string##1\rightbracket}}}
\def\writestoppt{}\def\writedef#1{}

\def\figin{\epsfcheck\figin}\def\figins{\epsfcheck\figins}
\def\epsfcheck{\ifx\epsfbox\UnDeFiNeD
\message{(NO epsf.tex, FIGURES WILL BE IGNORED)}
\gdef\figin##1{\vskip2in}\gdef\figins##1{\hskip.5in}
\else\message{(FIGURES WILL BE INCLUDED)}%
\gdef\figin##1{##1}\gdef\figins##1{##1}\fi}

\def\figinsert{}
\def\ifig#1#2#3{\xdef#1{fig.~\the\figno}
\writedef{#1\leftbracket fig.\noexpand~\the\figno}%
\figinsert\figin{\centerline{#3}}\medskip\centerline{\vbox{\baselineskip12pt
\advance\hsize by -1truein\center\footnotesize{  Fig.~\the\figno.} #2}}
\bigskip\endinsert\global\advance\figno by1}
\def\footnotefont{}\def\endinsert{}

\renewcommand{\thepage}{\arabic{page}}
\setcounter{page}{1}

\section{Introduction}

Surprisingly, very little is known about the behavior of
QCD at non-zero bare theta angle,
$\theta_0$. `t Hooft \cite{TH} argued that if QCD
continues to exhibit confinement at all $\theta_0$, then
a phase transition must occur at some finite $\theta_0$ -- most likely
$\theta_0 = \pi$. Indeed, this possibility is
supported by investigations using chiral Lagrangians \cite{CHPT}.
Alternatively, he suggested that the theory could
switch to a novel phase, such as one with `oblique confinement',
in some finite region of $\theta_0$ near $\pi$.
Recently, based on lattice investigations,
Schierholz \cite{SCHZ} has suggested that confinement breaks
down completely even for infinitesimal values of $\theta_0$,
leaving QCD in a Higgs phase.

Seiberg and Witten's \cite{SW,SW2} solution of low-energy $N=2$ SQCD,
combined with recent results extending the
exact solutions to models with explicit, soft-SUSY breaking
\cite{us,N2S},
provides a useful laboratory for the study of strong gauge
dynamics. The models in question exhibit electric confinement due
to condensation of magnetic monopoles, a mechanism
first proposed for QCD by `t Hooft and Mandelstam \cite{THM}.
In their picture the relevant low-energy degrees of freedom
are magnetic monopoles interacting via the long-range
fields of the Abelian subgroup $U(1)^2$ of $SU(3)$.
`t Hooft's argument \cite{TH}  that the nature
of confinement
could depend on the $\theta_0$ angle is a direct
consequence of the existence of magnetic excitations,
and the dependence of their effective charge on $\theta_0$
(the `Witten effect' \cite{WE}).

While it is still an open question whether QCD
behaves in the manner envisaged in \cite{THM},
the Seiberg-Witten model clearly has the correct
ingredients to further investigate `t Hooft's predictions:
it remains in the Coulomb phase at low energies
and exhibits light magnetic degrees of freedom.
In this paper we study the behavior of softly broken SQCD
models at non-zero $\theta_0$. We note that the highly non-trivial
consistency conditions on the Seiberg-Witten ansatz are as
equally valid at non-zero $\theta_0$ as at $\theta_0=0$.
It is necessary to break
supersymmetry completely to see any physical dependence on
$\theta_0$. In both the $N=2$ and $N=1$ models
studied in \cite{SW}, the gaugino remains massles and
hence any dependence on $\theta_0$ can be rotated away.
(See \cite{Konishi} for a discussion of $\theta_0$ and CP
invariance in this class of models.)
In the models that we study the anomalous $U(1)_R$
symmetry which may be used to generate shifts in $\theta_0$
is explicitly violated by SUSY breaking interactions such
as masses for the gauginos and adjoint quarks.
Hence the physics can and does depend on $\theta_0$ in
an interesting way.

\section{Review of  Solved Models}

In this section we review Seiberg and Witten's
solution of $N=2$ SQCD and the
spurion technique for studying the effects of
soft breaking perturbations. This paper
will concentrate on the case of an $SU(2)$
gauge group and a single $N=2$ vector multiplet
$A^a$. In terms of $N=1$ multiplets $A^a$ contains
a vector multiplet $( A^a_{\mu}, \lambda^a)$ and
a chiral multiplet $(\psi^a,\phi^a)$, all in the
adjoint representation.

\subsection{Pure Glue N=2 SQCD}

$N=2$ SUSY models are highly constrained and
can be characterized completely by a
prepotential ${\cal F} (A)$, which determines both
the Kahler- and super-potentials of the model.
At tree level the prepotential is given by ${\cal F}(A) = \frac{1}{2}
\tau_{cl} A^2$ where $\tau_{cl}$ contains the bare
gauge  coupling and $\theta_0$ angle
\beq
\label{tau}
\tau_{cl} = {\theta_0 \over 2 \pi} + i{4\pi \over g_0^2}
\eeq

The classical theory possesses a number of $U(1)_R$ symmetries
under which the superspace coordinate $\theta$ has charge +1,
$W_{\alpha}$ charge +1 and the matter fields arbitrary charge. In
the quantum theory the $U(1)_R$ symmetries are anomalous except
for that with matter field charge 0. The anomalous Ward identities
associated with the broken symmetries may be used to rotate $\theta_0$
onto the matter and gaugino mass terms and, since these are zero, to
rotate $\theta_0$ away without physical consequences.

Vacua of the classical theory are described by a moduli space with
the adjoint scalar vacuum expectation value (parameterized by the gauge
invariant
quantity $u= tr[a^2 ]$) as coordinates. At energy scales below $u$
the classical low energy theory contains only the $U(1)$ subgroup of the gauge
group (a photon and its gaugino) and the neutral components of the matter
fields.
At $u = 0$ the full $SU(2)$ symmetry is restored.
The low energy $U(1)$ theory may be written
in $N=1$ superspace notation in terms of $\tau_{cl}$ as
\beq
\label{Lag}
{\cal L} ~=~ \frac{1}{4 \pi} Im[ \int d^4 \theta
\frac{\partial {\cal F}}{\partial A} \bar{A} ~+~ \frac{1}{2}
\int d^2 \theta \frac{ \partial^2 {\cal F}} {\partial A^2} W_{\alpha}
W^{\alpha} ]~.
\eeq

Seiberg and Witten's ansatz for the solution of the quantum theory also
possesses
a moduli space in $u$ and is described below the
scale $u$ by a $U(1)$ theory with neutral
matter fields, ($
\psi, a$). In the quantum theory the $SU(2)$ symmetry is not restored
at $u=0$. The solution gives two descriptions of the low energy theory, one in
the
orginal electric variables and one in the electro-magnetic dual variables.
The explicit form of $a(u)$, $a_D(u)$ are given in
terms of the periods of a meromorphic differential of the second
kind on a genus one surface described by the equation:
\beq
\label{ellip}
y^2=(x^2-\Lambda^4)(x-u),
\eeq
describing the double covering of the plane branched at
${\pm \Lambda^2}$, $u$, $\infty$.
If one  chooses the cuts $\{-\Lambda^2, \Lambda^2\}$, $\{u, \infty\}$,
then the solution is represented by elliptic functions \cite{N2S}
\beq
\label{aad}
a(u)={4 \Lambda \over \pi k}E(k), \,\,\,\,\,\ a_D(u)={4 \Lambda \over
i \pi }{E'(k)-K'(k) \over k},
k^2={2 \over 1+u/\Lambda^2}.
\eeq
where
$$
K(k)={\pi \over 2}F(1/2, 1/2, 1;k^2); \,\,\ K'(k)=K(k');
$$
\beq
\label{xvii}
E(k)={\pi \over 2}F(-1/2, 1/2, 1;k^2); \,\,\ E'(k)=E(k'), \,\,\
{k'}^2+k^2=1,
\eeq
Note that $a_D = {\partial {\cal F} \over \partial a}$
and the  effective coupling constants $\tau$ is given by
\beq
\tau  =  {1 \over 2} {\partial^2 {\cal F} \over \partial a^2} =
{\partial a_{D} \over \partial a}={da_{D}/dk \over
da/dk}={iK' \over K}
\eeq

The scale $\Lambda$ generated by the theory
is related to the bare coupling constant $g_0$ and bare
$\theta_0$-parameter by
\beq
\label{Lam2}
\Lambda^2 ~=~ \Lambda_0^2 ~
{\rm exp}(-{4 \pi^2 \over g_0^2}+i{\theta_0 \over 2})
\eeq
This result is derived using the perturbatively exact
one-loop $\beta$ function which depends
holomorphically on the bare chiral superfield $\tau$.
An important consequence is the
dependence of $\Lambda$ on $\theta_0$.

In addition the above relations imply the following useful form for $u$
\cite{Matone}
\beq \label{wronk}
u = i \pi \left( {\cal F} - {1 \over 2} a { \partial {\cal F} \over \partial a}
\right)
\eeq

In fig. 1 we show the real and imaginary parts of the effective coupling $\tau$
as a function of $u$, which correspond to the effective theta
angle and the gauge coupling respectively.
The singularities in the $\tau$ function are consistent
with the hypothesis that a magnetic monopole (with electric and magnetic
charges $(0,1)$)
becomes massless at the point $+\Lambda^2$ and a $(-1,1)$
(or $(1,1)$, depending on whether one approaches from above
or below the branch cut) dyon becomes massless
at the point $-\Lambda^2$. Note that at the latter point the effective
angle, $\theta_{eff}$, is $2 \pi$ and hence by the Witten effect the dyon
behaves like a $(0,1)$ monopole at $\theta_{eff}=0$.
The theory in fact possesses a $Z_2$
symmetry that leaves the physics invariant under $u \rightarrow - u$.
The extra dyonic matter multiplets are
introduced into the effective theory close to the singular points.
Their interactions are constrained by $N=2$ SUSY to be of the form
\beq
\label{Lm}
{\cal L}_M=\int d^4 \theta \big( M^{\dagger}{\rm e}^{2V_D}M +
 {\widetilde M}^{\dagger}{\rm e}^{-2V_D}{\widetilde M}\big)+
 2{\sqrt 2} Re \int d^2 \theta A_DM {\widetilde M} ~.
\eeq
\vspace{1cm}

$\left. \right.$  \hspace{-2in}\ifig\prtbdiag{}
{\epsfxsize7truecm\epsfbox{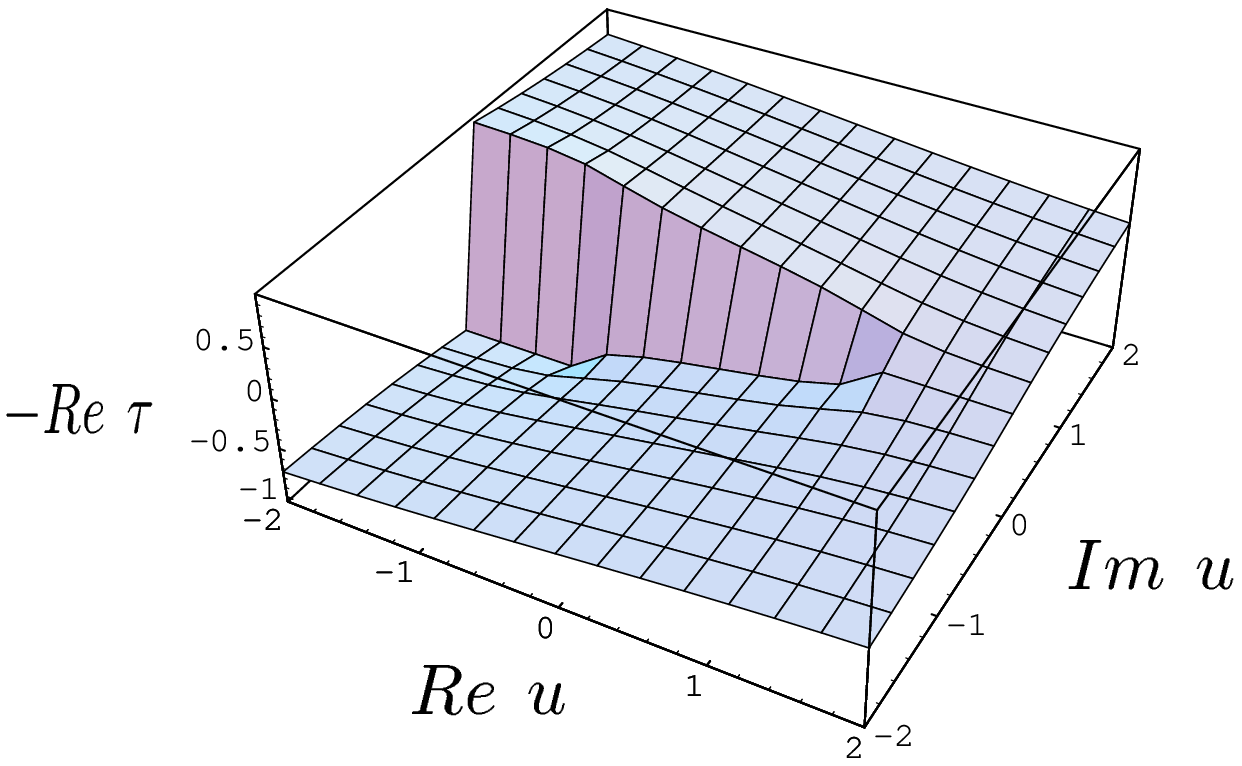}}  \vspace{-5.7cm} \\
$\left. \right.$ \hspace{4cm} \ifig\prtbdiag{}
{\epsfxsize7truecm\epsfbox{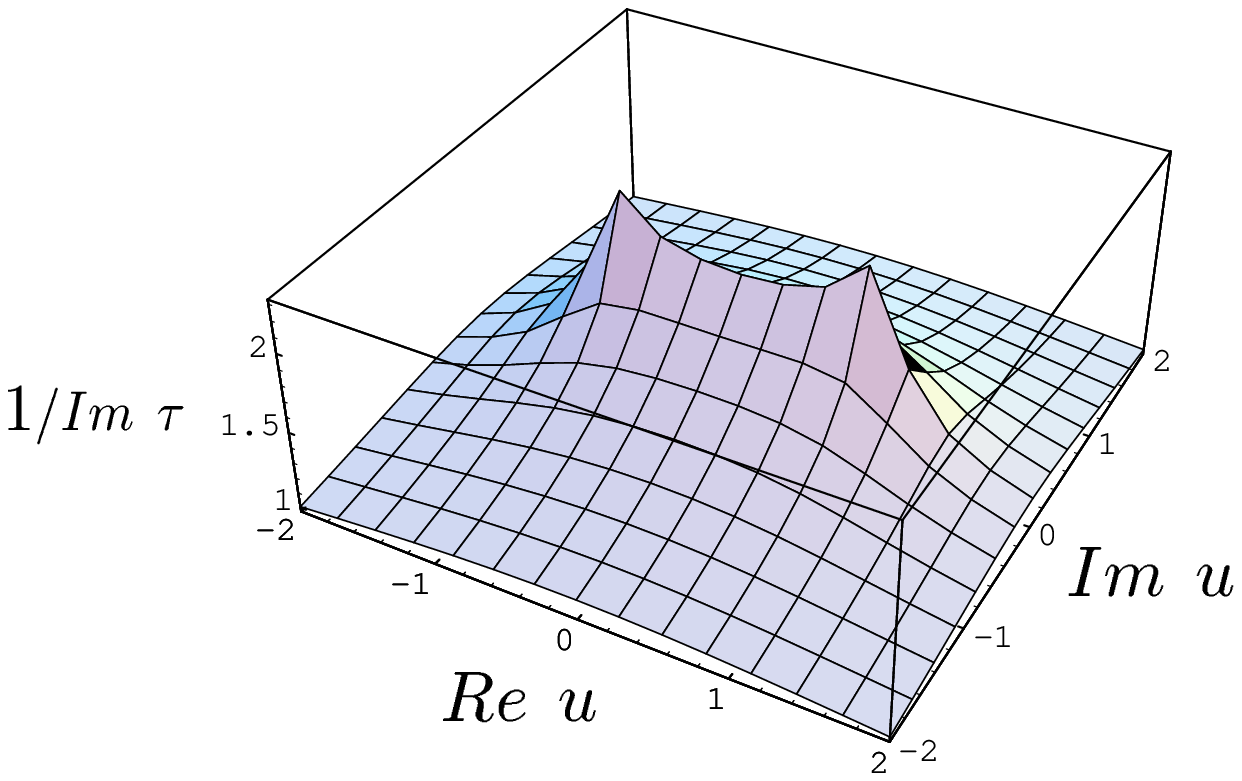}}\vspace{-1.6cm} \\

\begin{center} Fig.1: Plots of -$Re~\tau$ and $1/Im~\tau$
over the $u$ plane. \end{center}

The theory possesses an $SL(2,Z)$ symmetry acting on $(a_D, a)$ and
hence also on $(\tau,1)$. The two generators of the transformations are
\beq
\label{Tdual}
T = \left( \begin{array}{cc} 1 & 1 \\ 0 & 1 \end{array} \right)
\eeq
which corresponds to shifting  $\theta_{eff}$  by $2 \pi$, and
\beq
\label{Sdual}
S = \left( \begin{array}{cc} 0& 1 \\ -1 & 0 \end{array} \right)
\eeq
which corresponds to transforming to the electromagnetic dual variables.
There are thus a number of different descriptions of the theory at any
particular value of $u$. We shall make use of the following descriptions:
at large $u$ the orginal electric variables in (\ref{aad}) are the weakly
coupled
description (we shall use superscript $w$); near the singularity at $+
\Lambda^2$
with a light monopole field the electric variables are strongly coupled but the
magnetic variables are weakly coupled (superscript $m$); and near the
singularity
$- \Lambda^2$ the magnetic variables are again weakly coupled but in order
to make the couplings of the $(-1,1)$ (or $(1,1)$) dyon to the photon
local we shift $\theta_{eff}$ by $2 \pi$ using the generator $T$ (these
variables
have superscript $d$).

\subsection{$N=1$ Solution}

The $N=2$ solution can be perturbed by the addition of a small $N=1$
preserving mass for the adjoint matter field \cite{SW}
\beq
\label{MASS}
{\cal L}_{pert} = 2  Re \int d^2 \theta ~ m U(A)
\eeq

The $N=1$ theory exhibits a classical $U(1)_R$ symmetry
with charges $\theta$ +1, $W_\alpha$ +1 and $U$ +2. It is
anomalous in the quantum theory and can be used to rotate
$\theta_0$ onto the gaugino mass term and hence eliminate it
from the theory.

The dependence of the effective theory's superpotential on the bare mass term
is that of the bare theory up to gauge renormalization.
The minima of the potential are given
by the extrema of the superpotential and are hence determined
by
\begin{eqnarray}
\sqrt{2} a_M a_{\tilde{M}} + m \kappa & = & 0  \\
a^{(m)} a_M = a^{(m)} a_{\tilde{M}} & = & 0
\end{eqnarray}
where
\beq \kappa = {\partial u \over \partial a^{(m)}} = i \pi \left(
a_D^{(m)} - {1 \over 2} a^{(m)} \tau^{(m)} \right)
\eeq
and we have used (\ref{wronk}). We find
\begin{eqnarray} \label{N=1cond}
a_D & = & 0 \\
a_M = a_{\tilde{M}} & = & \left( -m\kappa(0)/\sqrt{2} \right)^{1/2}
\end{eqnarray}
The vacuum energy is now minimized at the singular points where there is a
condensate of the magnetic monopole field and $\theta_{eff} =0$.
In terms of the orginal electric variables there is confinement
by the `t Hooft--Mandelstam mechanism. We note that since confinement sets in
on
the scale of the $N=2$ breaking we must have $m \ll \Lambda$ to retain a $U(1)$
phase
and hence the validity of the effective theory.

\subsection{Softly Broken $N=2$ SQCD}

Soft SUSY breaking interactions may be introduced into the exact solution of
$N=2$ SQCD through the $F$-components of the `spurion' coupling fields. We
treat
the couplings $\tau$ and $m$ as the vevs of the lowest components of chiral
superfields
and their occurence in the effective theory is hence constrained by
SUSY and the spurions' $U(1)$ couplings.
The requirement that the theory returns to the SUSY limit as the $F$-components
are turned off constrains how the couplings may
occur in the theory up to possible D-terms that
vanish in the SUSY limit which we shall discuss below where appropriate.

\subsubsection{$N=1$ Spurions}

A first example of a solvable (at order $p^0$) softly broken model is to start
from the $N=1$ solution described above and allow the spurion field $m$
to acquire a non-zero F-component vev \cite{us}. At tree level this induces
the additional $N=1$ violating interaction
\beq
\label{fm}
2 Re ( f_m u)~,
\eeq
This term explicitly breaks the $Z_2$
and anomalous $U(1)_R$ symmetries of the supersymmetric
$N=1,2$ models. There is no anomalous $U(1)$ symmetry in
the theory and hence $\theta_0$ can have physical consequences.

By requiring the $N=1$ limit as $f_m \rightarrow 0$ we observe that the
superpotential is not renormalized in the quantum theory. Treating
$m$ as a spurion field there is a non-anomalous $U(1)_R$
symmetry under which $m$ transforms with charge $+2$
and hence may only occur in the D-terms as $m^\dagger m /\Lambda^2$.
These unknown terms are an order down in the momentum
expansion (we take $p \sim m << \Lambda$)
compared to the soft breaking terms from the
superpotenial.
Hence (\ref{fm}) is the leading order correction
to the low-energy effective potential
when $f_m / \Lambda^2 << a_m / \Lambda << 1$.
We  obtain the following potential for
the softly-broken model
\beq
V = 2 |a^{(m)}|^2 ( |a_M|^2 + |a_{\tilde{M}}|^2)  + {1 \over 2 b}
( |a_M|^2 + |a_{\tilde{M}}|^2)^2 + {1 \over b} ( |\kappa m|^2 +
2\sqrt{2} Re(a_M a_{\tilde{M}} \kappa^* m^*) ) - 2 Re(f_m u)
\eeq
with
\beq
b = {1 \over 4 \pi} Im \tau^{(m)} , \hspace{1cm} \kappa = {\partial u \over
\partial a^{(m)}}
\eeq
Minimizing the potential with respect to the monopole fields yields
$|a_M|^2 = |a_{\tilde{M}}|^2$. We have
\beq
a_M = i \rho e^{i \alpha}, \hspace{1cm} a_{\tilde{M}} = i \rho e^{i \beta}
\eeq
where for a minimum we have $\alpha + \beta = {\rm phase} (m \kappa)$.
{}From which we find
\beq
\label{Vmin1}
V=-{2 \over b} \rho^4 + {| \kappa m|^2 \over b} - 2 Re f_m u
\eeq
with the  monopole condensate
\beq
\label{N1cond}
\rho^2= \cases{ -b|a^{(m)}|^2 + { 1 \over {\sqrt 2} }  |\kappa m| >0 \cr
0 }~~.
\eeq
To obtain the potential in the dyon region we note  that only the final
term in (\ref{Vmin1}) violates the $Z_2$
symmetry of the $N=1$ limit.
In the dyon region we therefore have
\begin{eqnarray}
 V^{(d)}(u)  & = & V^{(m)}(-u)  - 4 Re f_m u \\
\rho^2_{(d)}(u) & = & \rho^2_{(m)}(-u)
\end{eqnarray}
The potential may now be minimized numerically. The potential is shown for
$\theta_0=0$, $m/\Lambda = 0.01$ and $f_m/\Lambda = 0.001$ as
a function of $u$ in fig. 2.  The $Z_2$ symmetry is explicitly
broken and the global minimum lies close to the point $+ \Lambda^2$
as we expect since we are only perturbing the $N=1$ model. By choosing
$f_m$ to be complex it is possible to perturb the minima away from
the singularity in any direction on the $u$ plane.  Since the singularity lies
on the cut it is possible to vary $f_m$ smoothly and see a discontinuous
jump in the value of $\theta_{eff}$ as the minima crosses the cut.

\vspace{-1.5cm}

$\left. \right.$  \hspace{-2cm}\ifig\prtbdiag{}
{\epsfxsize8truecm\epsfbox{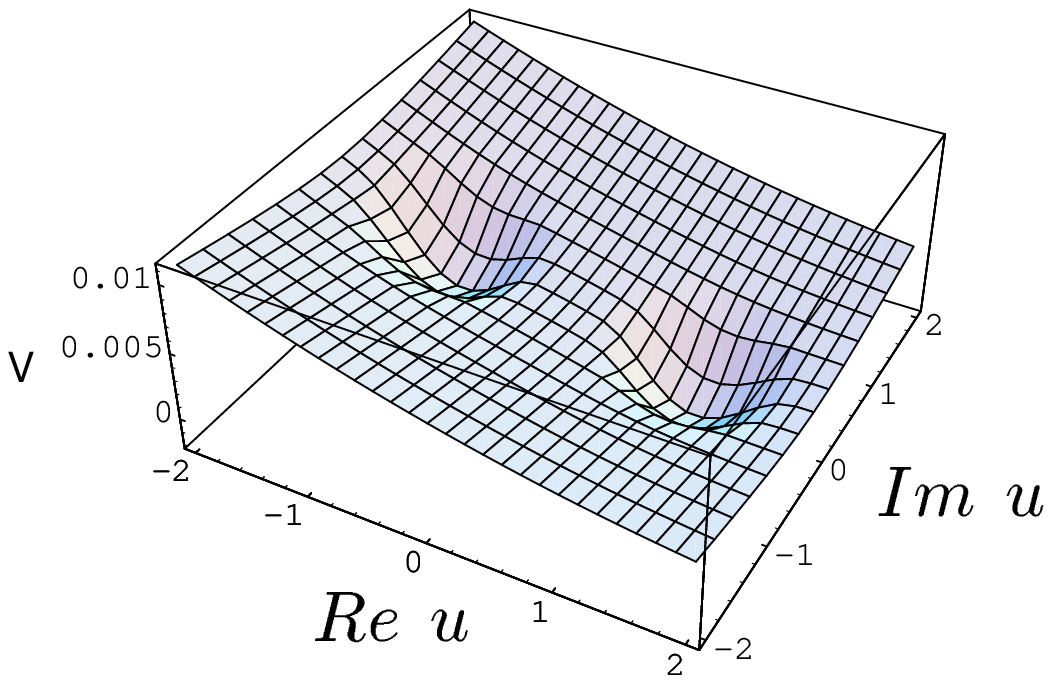}}  \vspace{-2cm} \\

\begin{center} Fig.2: Potential in $N=1$ spurion model as a function of $u$
for $m=0.01$ and $f_m=0.001$. \end{center}

\subsubsection{$N=2$ Spurion}

Another method for introducing soft breakings in
a controllable way is to consider a full $N=2$ vector
multiplet of spurion fields \cite{N2S}.
This has the advantage of retaining control over the
D-terms, although it produces a very restricted set
of breaking terms. Following \cite{N2S},  let us introduce the spurion field
$S$ as a second vector multiplet
\beq
{\cal L} ~=~ \frac{1}{4 \pi} Im \left[ \int d^4 \theta  \left(
\frac{\partial F}{\partial A} \bar{A} ~+~
\frac{\partial F}{\partial S} \bar{S} \right) + \frac{1}{2}
\int d^2 \theta \left( \frac{ \partial^2 F} {\partial A^2} W_{\alpha}
W^{\alpha}
+
 \frac{ \partial^2 F} {\partial A \partial S} W_{\alpha} W^{'\alpha}
+
 \frac{ \partial^2 F} {\partial S^2} W_{\alpha}' W^{'\alpha} \right) \right]~.
\eeq
with bare prepotential
\beq
{\cal F}_{cl} = {1 \over 2 \pi} S A^2
\eeq
For convenience we define the $2\times 2$ matrix of couplings:
\beq
\label{3tau}
\tau_{11}={\partial^2 {\cal F} \over \partial^2 a},\,\,\,\
\tau_{01}={\partial^2 {\cal F} \over \partial s \partial a},\,\,\,\
\tau_{00}={\partial^2 {\cal F} \over \partial ^2 s}~.
\eeq
Freezing  the scalar component of the second
multiplet's matter field generates the coupling
of the pure glue model with $s_{cl} = \pi \tau_{cl}$ (this normalization
is convenient since $\Lambda \sim exp(is)$).
We  may also freeze the F-component of the spurion matter field,
$F_0$, and generate soft breaking interactions.
At the bare level, this induces the following SUSY
violating interactions in the $SU(2)$ model:
\beq \label{bare}
{\cal L}_{SSB} ~=~  {1 \over 8 \pi^2} Im ( F_0^*  \psi_A^\alpha  \psi_A^\alpha
+ F_0 \lambda^\alpha \lambda^\alpha) -
{|F_0|^2 \over  4 \pi Im \tau_{cl}} (Im~a^\alpha)^2
\eeq
The gauginos and matter fermions acquire masses and
hence it is again not possible to rotate away $\theta_0$.
The imaginary components of the scalar fields also acquire
a mass at tree level. Since SUSY is explicitly broken in the theory
and there are thus no symmetries protecting the scalar masses
there will be quadratic divergences in the theory below the
SUSY breaking scale that will radiatively generate masses
for the real components of the scalar fields as well.  Thus we expect
that if $F_0$ were taken to infinity we would recover pure glue
(non-SUSY) QCD.

The soft breaking terms again break the $Z_2$ symmetry
of the pure N=2 model.
At the quantum level, the effects of the SUSY breaking
on the effective theory can be computed using the
holomorphic dependence of the prepotential
${\cal F} (A)$ on the spurion field; that is the quantum prepotential
is just that of the pure glue theory.

In the weakly coupled region (far from the singularities) the theory is best
described in terms of the electric variables  \cite{N2S} :
$$
a_D^{(w)}={4 \Lambda \over i\pi}{E'-K' \over k},
\,\,\,\,\ a^{(w)}={4 \Lambda \over \pi k}E(k),
$$
\beq
\label{HGS}
\tau^{(w)}_{11}={iK' \over K},\,\,\,\
\tau^{(w)}_{01}={2\Lambda \over kK},\,\,\,\
\tau^{(w)}_{00}=
-{8i \Lambda^2\over \pi} \Big( {E-K \over k^2K} + {1 \over 2}\Big)~.
\eeq

In the region close to the singularity associated with a massless monopole
the dual variables are weakly coupled.
Using (\ref{HGS}) in combination with $SL(2,Z)$ and the residual
$Z_2$ symmetry, one can derive
expressions for the couplings in the monopole region

$$
a^{(m)}=a^{(w)}_D ,\,\,\,\,\
a_D^{(m)}= - a^{(w)},
$$
\beq
\label{MON}
\tau^{(m)}_{11}= - {1 \over  \tau^{(w)}_{11}},\,\,\,\
\tau^{(m)}_{01}= i \tau^{(w)}_{01},\,\,\,\
\tau^{(m)}_{00}={8i \Lambda^2\over \pi}
\Big( {E'\over k^2K'} -{1 \over 2}\Big).
\eeq
\noindent and  in the dyon  region
$$
a^{(d)}(u)=i \left(a_D^{(m)}(-u)-a^{(m)}(-u)\right),\,\,\,\
a_D^{(d)}(u)=i a^{(m)}(-u),
$$
\beq
\label{DYO}
\tau_{11}^{(d)}(u)=\tau_{11}^{(m)}(-u)-1,\,\,\,\
\tau_{01}^{(d)}(u)=i \tau_{01}^{(m)}(-u),\,\,\,\
\tau_{00}^{(d)}(u)=-\tau_{00}^{(m)}(-u)~~.
\eeq
Note that in the dyon region we have made an
$SL (2, Z)$ transformation to shift
$\theta_{eff}$  by $2 \pi$, so that the dyon has
a local coupling to the photon, with charge $(0,1)$.
$\tau^{(d)} (u)$ is then written in terms of
$\tau^{(m)} (-u)$ in order to avoid the branch cut to infinity when
$u$ is in the dyon region $u \sim -\Lambda^2$.
The functions $a^{(m)}$ and $( a^{(d)} + a_D^{(d)} )$ disappear
at the corresponding singular points, leading
to massless monopoles and dyons.

The potential  is obtained by eliminating the auxiliary fields $F_M$,
 $F_{\widetilde M}$ and $F_a$ :
\beqa
\label{V}
V &=& {1 \over 2b_{11}}\big( |a_M|^2+| a_{\widetilde M}|^2 \big)^2 + 2|a|^2
\big(
|a_M|^2+|a_{\widetilde M}|^2 \big) \nonumber \\
& & \nonumber\\
 &+& {1 \over b_{11}} {\sqrt 2}b_{01}
\big({\overline F_0}a_M a_{\widetilde M} + F_0{\overline a_M}
{\overline{a_{\widetilde M}}} \big) \nonumber - {{\rm det} \,\ b_{ij} \over
b_{11}}
|F_0|^2,
\eeqa
where
\beq
b_{ij}\equiv {1 \over 4 \pi}{\rm Im} \,\ \tau_{ij} =
{1 \over 4\pi}{\rm Im} {\partial^2 {\cal F} \over \partial a^i \partial a^j}.
\eeq
$a_M$, $a_{\widetilde M}$ are, as before, the scalar components of the
(monopole or dyon ) superfield, $M$,  and all fields are taken in the
corresponding patches.

Minimized with respect to the monopole (dyon) fields it becomes
(here $f_0 = \vert F_0 \vert$ and is real)
\beq
\label{Vmin2}
V=-{2 \over b_{11}} \rho^4-{{\rm det}b \over b_{11}}f_0^2
\eeq
with the  monopole(dyon) condensate
\beq
\label{cond}
\rho^2= \cases{ -b_{11}|a|^2+{1\over {\sqrt 2}}|b_{01}|f_0 >0 \cr
0 } ~~.
\eeq

The resulting potential is plotted in fig. 3 for $\theta_0 =0$. The soft  SUSY
terms  lift the degeneracy of the moduli space,
and a unique vacuum appears near  $ + \Lambda^2$. The $Z_2$ symmetry
is explicitly broken as can be seen in fig. 3. It is also explicitly broken by
the regions of  the $u$ plane in which the monopoles and dyons acquire
vevs which we show in fig. 4. As
$F_0$ increases the minima moves outwards along the real $u$
axis where  $\theta_{eff} = 0$. \vspace{0.5cm}

$\left. \right.$  \hspace{-2cm}\ifig\prtbdiag{}
{\epsfxsize8truecm\epsfbox{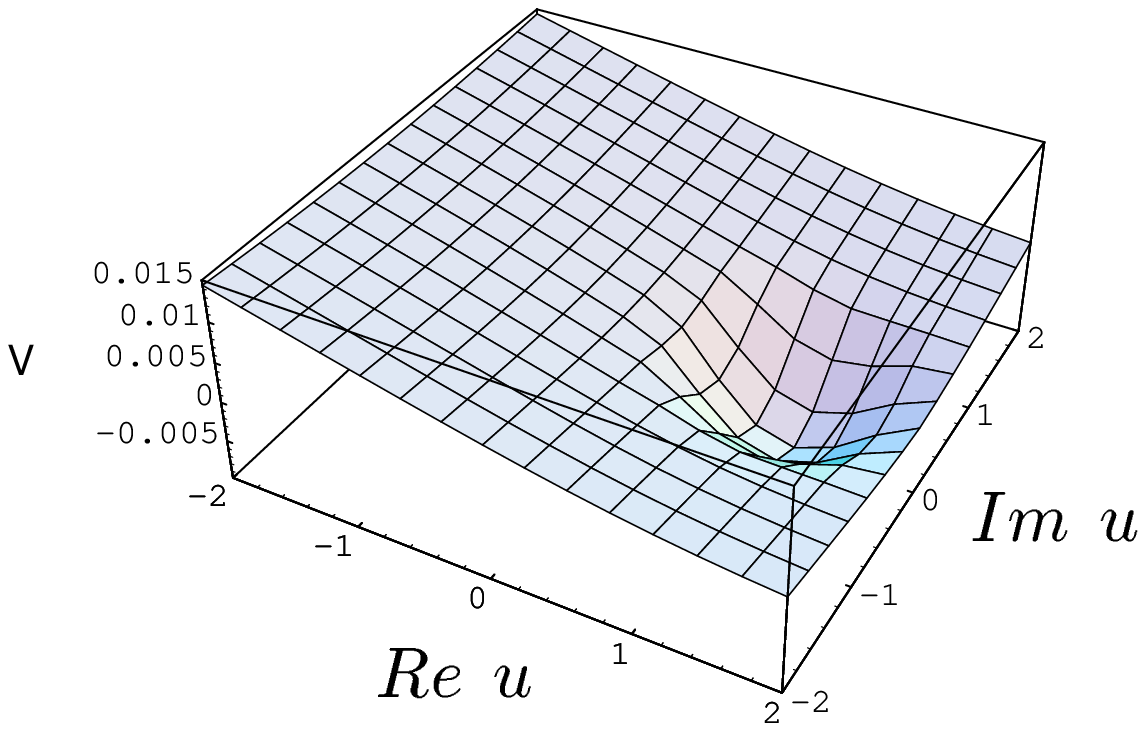}}  \vspace{-2cm} \\

\begin{center} Fig.3: Potential for $N=2$ spurion with $\theta_0 =0$,
$F_0/\Lambda = 0.3$ . \end{center} \vspace{0.5cm}

\vskip 1cm

$\left. \right.$  \hspace{-5cm}\ifig\prtbdiag{}
{\epsfxsize6.5truecm\epsfbox{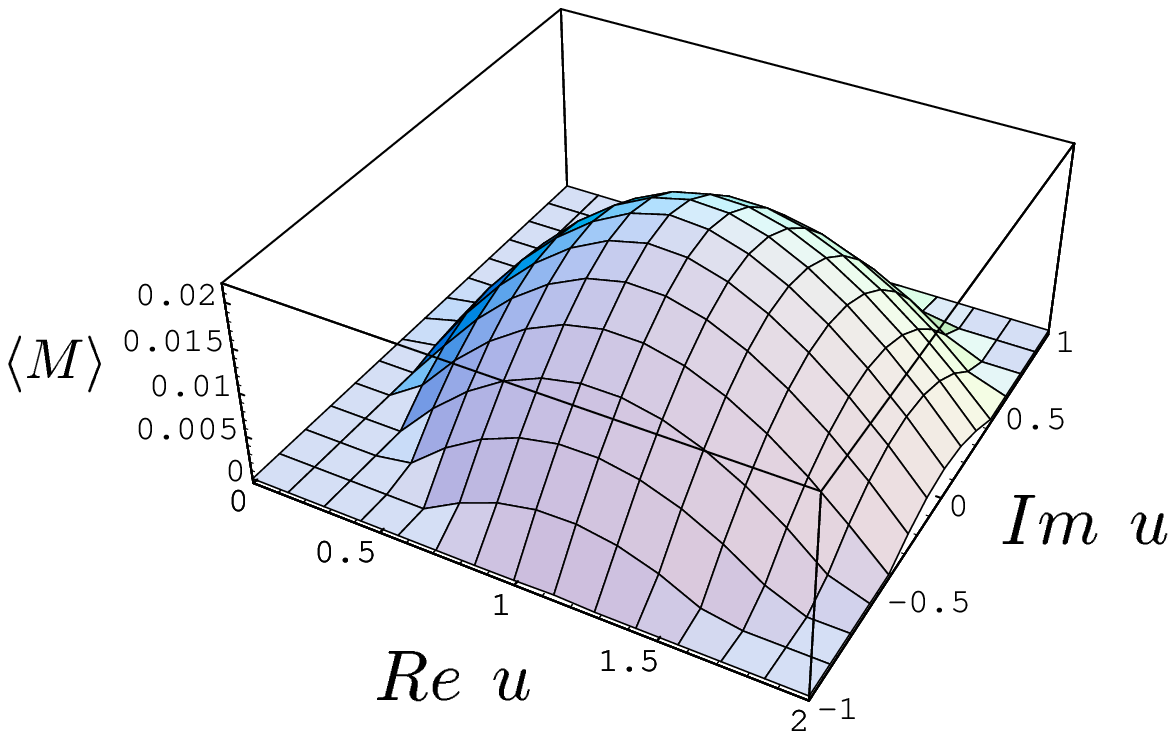}}  \vspace{-5.6cm} \\
$\left. \right.$  \hspace{3.5cm}\ifig\prtbdiag{}
{\epsfxsize7truecm\epsfbox{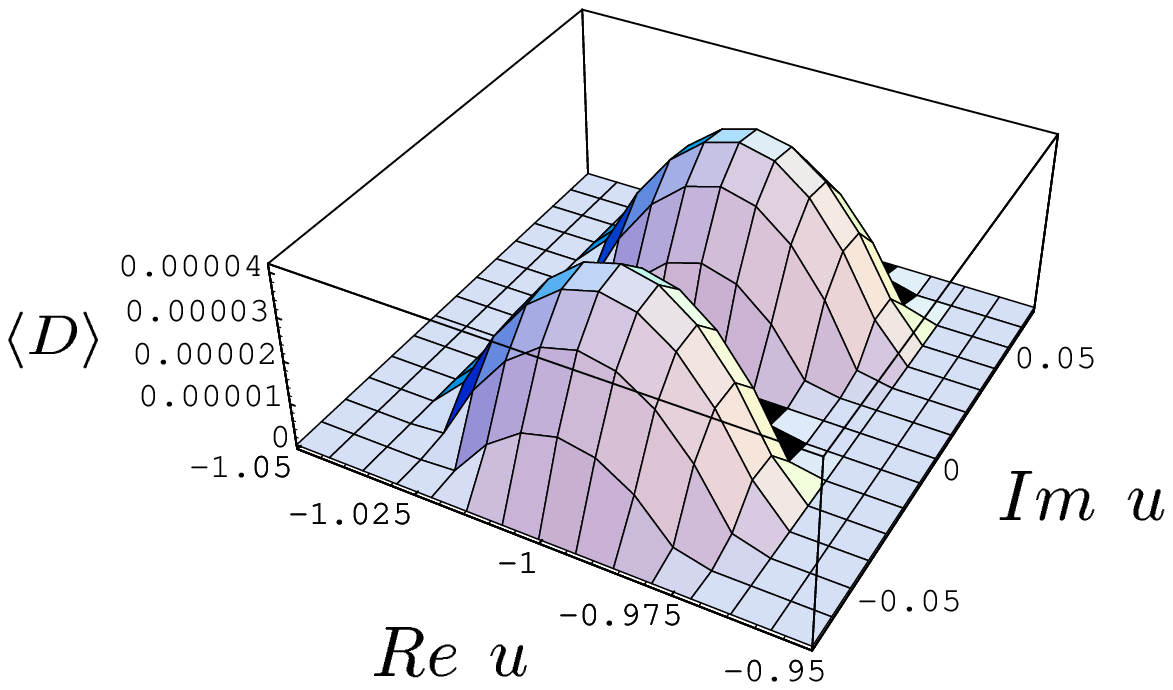}}  \vspace{-2cm} \\

\begin{center} Fig.4: Regions of the $u$ plane with monopole, $\langle M
\rangle$,
and  dyon condensates, $\langle D \rangle$, in the $N=2$ spurion model at
$\theta_0 =0$,
$F_0/\Lambda = 0.3$. \end{center}

\vskip 1cm

\section{Dependence On The Bare $\theta_0$ Angle}

We now turn to our discussion of the dependence of the solutions
described above on the bare  theta angle, $\theta_0$. `t Hooft argued in
\cite{TH}
that theories that confine by monopole condensation must exhibit a phase
transition as  $\theta_0$  crosses $\pi$ and conjectured that new phases
such as `oblique' confinement (confinement due to the condensation
of a purely magnetically charged mode which is the bound state of
dyons of opposite electric charge)
might occur in a patch around $\pi$. His argument
is essentially that due to the `Witten effect' as $\theta_0$ changes from 0 to
$2 \pi$
the charge of the (0,1) monopole causing confinement at $\theta_0 =0$ must
become (1,1). Since the theory must be $2 \pi$ periodic in $\theta_0$ a  dyon
in the theory must become the (0,1) state at $\theta_0 = 2 \pi$. As $\theta_0$
changes the vacuum must switch between having condensates of these
two different dyons so one expects a first order phase transition, most likely
at
$\theta_0 = \pi$.

Phase transitions in the effective chiral lagrangian of QCD with non-zero
$\theta_0$
have also been observed \cite{CHPT}.
Here $\theta_0$ can be rotated onto the quark mass
matrix as a phase $\theta_0/N_f$ where $N_f$ is the number of quark
flavors.  The theory is again $2 \pi$ periodic in $\theta_0$ but the effective
theory
depends on $\theta_0/N_f$ which leads to $N_f$ distinct, degenerate vacua.
Phase
transitions between these vacua occur at
$\theta = ( {\rm odd  ~integer} ) \cdot \pi$.

The exact solutions of
N=2 SQCD models allow further investigations of
the $\theta_0$ dependence of confinement since they confine by the mechanism
envisioned by `t Hooft. However the behavior of the
SUSY theories is more
subtle than the basic
argument suggested above since  $\theta_0$ is renormalized and
leads to different effective theta angles, $\theta_{eff} (u)$, at different
points on the
$N=2$ theory's moduli
space.

The effective theories depend on $\theta_0$ through the dynamically
generated scale $\Lambda$
\beq
\Lambda^2 \sim  exp (i \theta_0/2)~.
\eeq
The effect of changing $\theta_0$ in the pure glue $N=2$ theory is therefore
to rotate the positions of the singularities on the moduli space by an angle
of $\theta_0/2$.  For every point on the moduli space at one value
of $\theta_0$ there is an equivalent point on the moduli space at any other
value of $\theta_0$. The theory is therefore unchanged by changing $\theta_0$.
This is as expected since the $N=2$ theory has an anomalous $U(1)_R$ symmetry
acting on the gauginos and fermionic matter field
which are exactly massless. The anomalous Ward identity
may be used
to rotate $\theta_0$ away.

When the theory is perturbed to an $N=1$ theory by adding
a mass term for the scalar and fermionic matter fields
as described in section 2.2 the theory
is pinned at the singular points on the moduli
space at $\pm \Lambda^2$. The analysis
in section 2.2 is independent of the
phase of  $\Lambda$ as pointed out in \cite{Konishi}
and hence the theory is
pinned at the singularities for all $\theta_0$. The
effective $\theta_{eff}$ angle at the singularities
is 0 and $2 \pi$ respectively at the points with
 (0,1) and (-1,1) massless dyons.  Thus $\theta_{eff} = 0$
for any $\theta_0$. Again this is the expected
result since the gauginos of the theory are still massless
and the anomalous $U(1)_R$ symmetry may be used to rotate away $\theta_0$.

The problem becomes more interesting when the theory is softly
broken to N=0 since the ability to rotate away $\theta_0$ is lost.
In the case of the $N=1$ spurion there is no anomalous $U(1)$
symmetry and in the case of the $N=2$ spurion all the fermions
in the model obtain masses.
In addition the $Z_2$ symmetry of the effective theory
is broken. Using the potentials given in section 3.2,
we can track the groundstate of the system
as $\theta_0$  is varied. In addition to looking for the
phase transition predicted by 't Hooft (or the disappearance of
confinement above some critical $\theta_0$ -- which
we do {\it not} observe),
we are particularly interested
in the value of $\theta_{eff}$ at the minimum of the potential,
as this determines the effective charge of the dyon/monopole
condensate. Since the $N=1$ minima at $\pm \Lambda^2$ are directly
on the singular points in $\tau$, there is the possibility of discontinuous
behavior in $\theta_{eff}$ as the soft breakings force the $N=0$
minima away from the singularities.

\subsection{$\theta_0$ Dependence for the N=1 Spurion}

The potential of section 2.3.1 may be used to plot the potential of the
$N=1$ spurion model with changing $\theta_0$. For real $f_m$ and $m$
we show the results ( $m/\Lambda=0.01$ and $f_m/\Lambda=0.001$ in fig. 5.

At $\theta_0 = 0$ the minima close to $ + \Lambda^2$  lies below
that at $ - \Lambda^2$. As $\theta_0$ is raised towards $\pi$ the
line connecting the singularities
rotates in the $u$ plane by $\theta_0/2$
and the energy
difference between the global and local minima decreases. At $\theta = \pi$
the two vacua at $u \sim \pm i \Lambda^2$ are degenerate in energy. For
$\theta_0 > \pi$ the vacua at the singularity with the (-1,1) dyon becomes
the new minima of the theory. There is thus a first order phase transition
at $\theta_0 = \pi$ where distinct vacua interchange.

The degeneracy of the two vacua at $\theta_0 = \pi$ may be easily
seen from  (\ref{Vmin1}) and (\ref{N1cond}). The potential is that of
the $N=1$ model which possesses a reflection symmetry about the axis
defined by $exp(i\pi/2 + i\theta_0/2)$ plus the final term proportional
to $f_m$ which possesses the symmetry $u \rightarrow u^*$. At $\theta_0 =
\pi$ these two symmetries are identical and the whole
theory acquires a symmetry $u \rightarrow u^*$. Thus, provided the minima are
away from the origin, there will be two degenerate minima.
The phase transition is marked not only by a discontinuous jump in
$u$ but also in $\theta_{eff}$. We show this discontinuity, which grows with
$f_m$, in fig. 6.


$\left. \right.$  \hspace{-1cm}\ifig\prtbdiag{}
{\epsfxsize16truecm\epsfbox{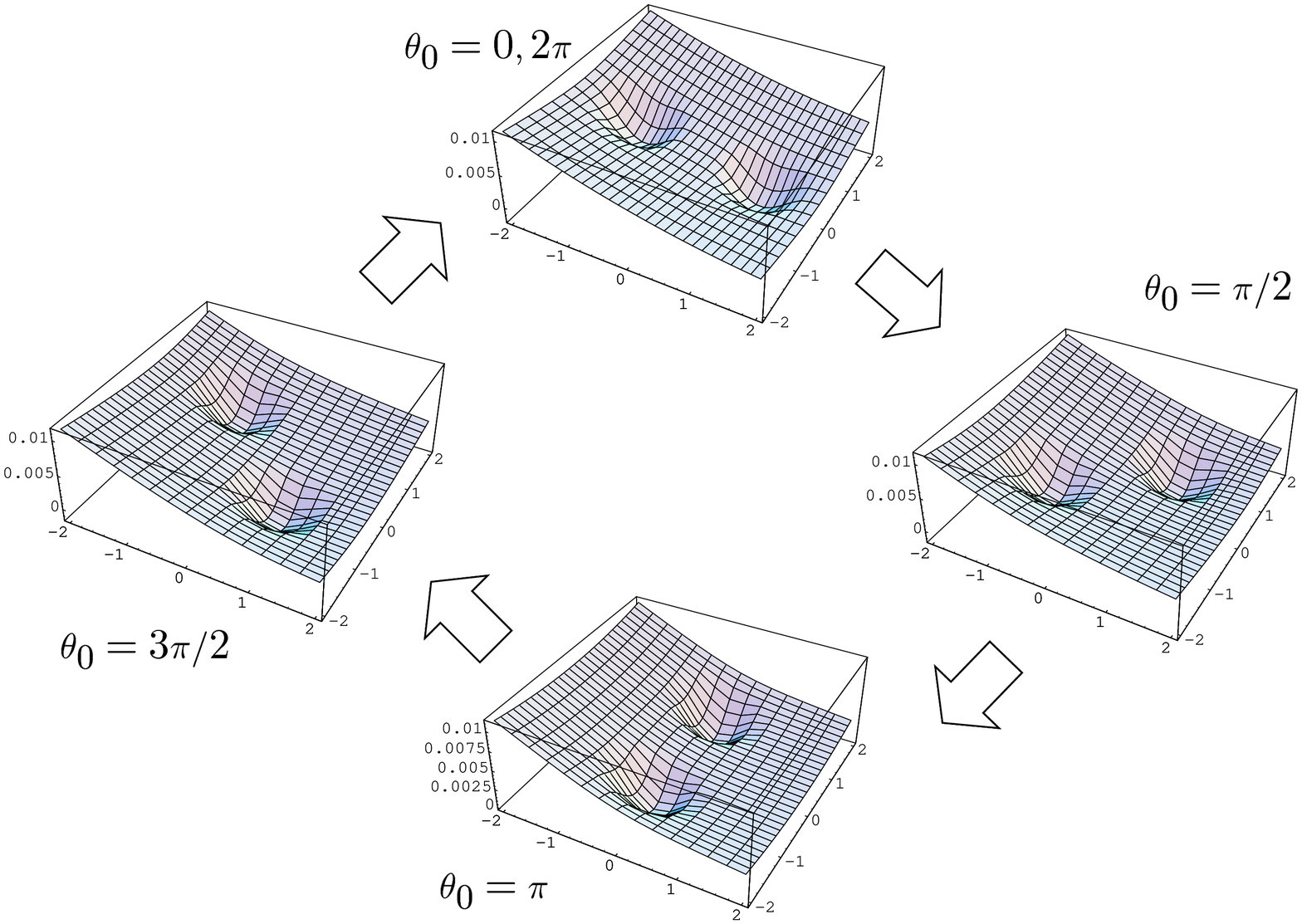}}  \vspace{-1cm}

\begin{center} Fig.5: Movement of minima with changing $\theta_0$
in the $N=1$ spurion model. \end{center}

\vspace{2.5cm}

$\left. \right.$  \hspace{-2cm}\ifig\prtbdiag{}
{\epsfxsize14truecm\epsfbox{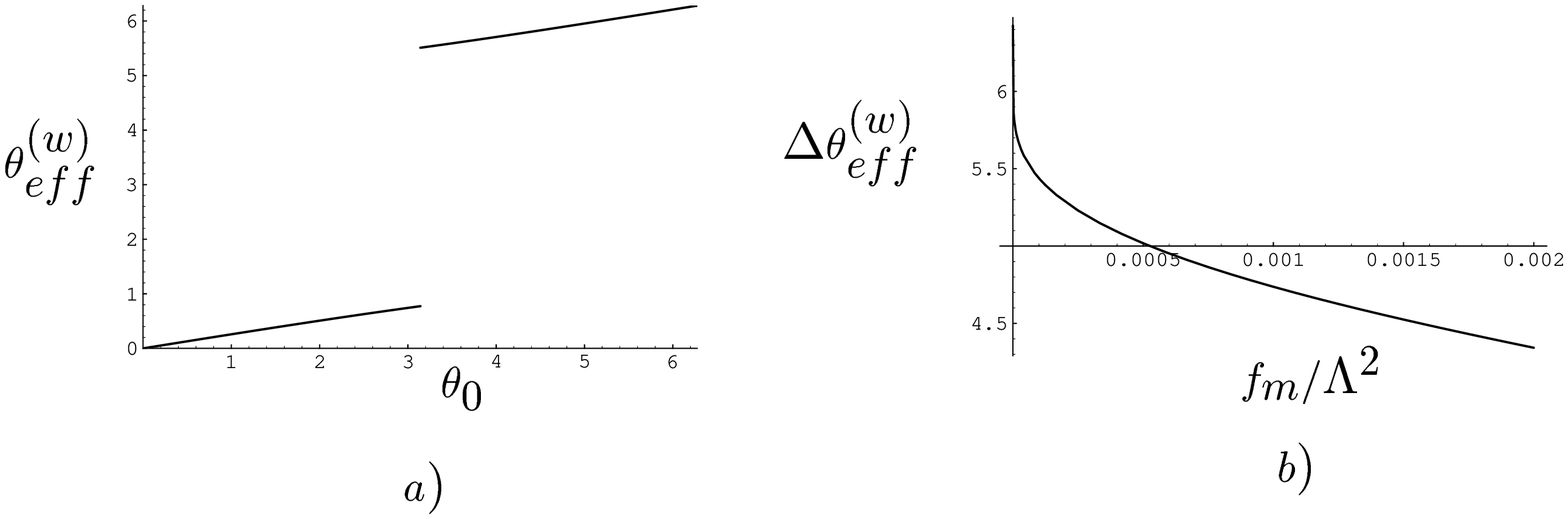}}  \vspace{-2cm} \\

\begin{center} Fig.6:  a) $\theta^{(w)}_{eff}$ as a function of 
$\theta_0$ at the
minima of the potential
in the $N=1$ spurion model, b) the discontinuity in
$\theta^{(w)}_{eff}$ at the phase transition
as a function of the soft breaking $f_m$. \end{center}

Note that in the above discussion we have taken
$f_m$ real. In the case of $f_m$ complex the phase transition
can occur at $\theta_0 \neq \pi$. For example, when $\theta_0$ is
purely imaginary the minima are exactly degenerate at $\theta_0 = 0$,
and the phase transition occurs as $\theta_0$ passes through
zero.

\subsection{$\theta_0$ Dependence for the $N=2$ Spurion}

For the N=2 spurion we may plot the potential
of section 2.3.2 for varying $\theta_0$.
The effective theory again has a first order phase transition
at $\theta_{eff} = \pi$. We demonstrate this in fig. 7
by
 showing the potential for varying $\theta_0$.
Again, the transition is a result of the breaking
of the
$Z_2$ symmetry of the model and the fact that the
effect of changing $\theta_0$ is effectively to rotate the $u$ plane by
$\theta_0 /2$. The bare theory is $2 \pi$ periodic in $\theta_0$
so the singularity with a massless (1,1) dyon (at $\theta_0 =0$)
must become the new minima of the theory at $\theta = 2 \pi$.
It is interesting to plot the region
in which there is a  monopole condensate for varying $\theta_0$
since the single region at $\theta_0 =0$ must transmute to the two regions of
condensation, observed for the dyon at $\theta_0 =0$, as $\theta_0$
changes to $2 \pi$. We show this transformation in fig. 8.

\vskip .5 in

$\left. \right.$  \hspace{-1cm}\ifig\prtbdiag{}
{\epsfxsize14truecm\epsfbox{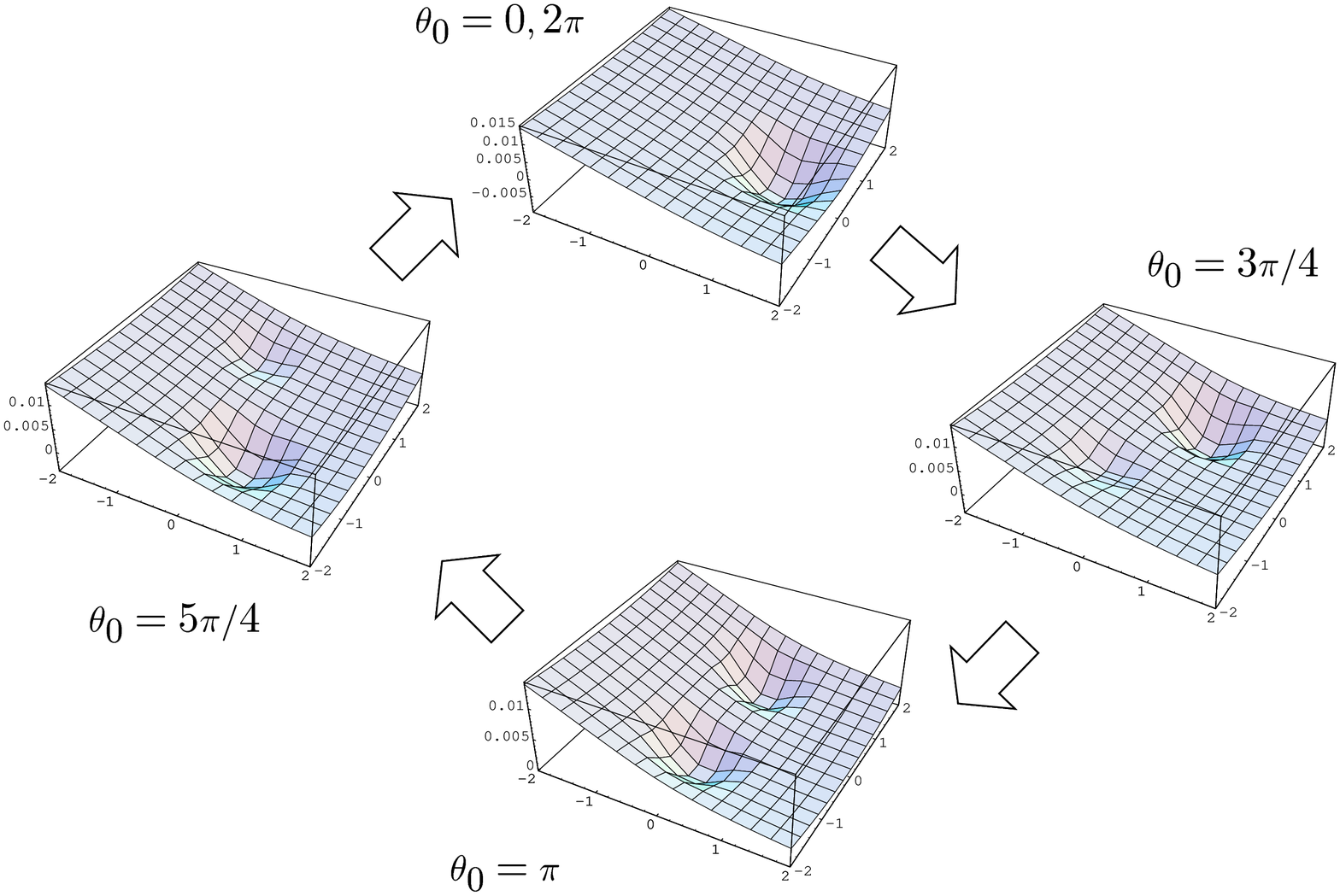}}  \vspace{-1cm}

\begin{center} Fig.7: Potential for N=2 spurion with $\theta_0 =0, \pi/2, \pi,
3 \pi/2$ and $2 \pi$, $F_0/\Lambda = 0.3$, showing the first order phase
transition. \end{center}

$\left. \right.$  \hspace{-1cm}\ifig\prtbdiag{}
{\epsfxsize16truecm\epsfbox{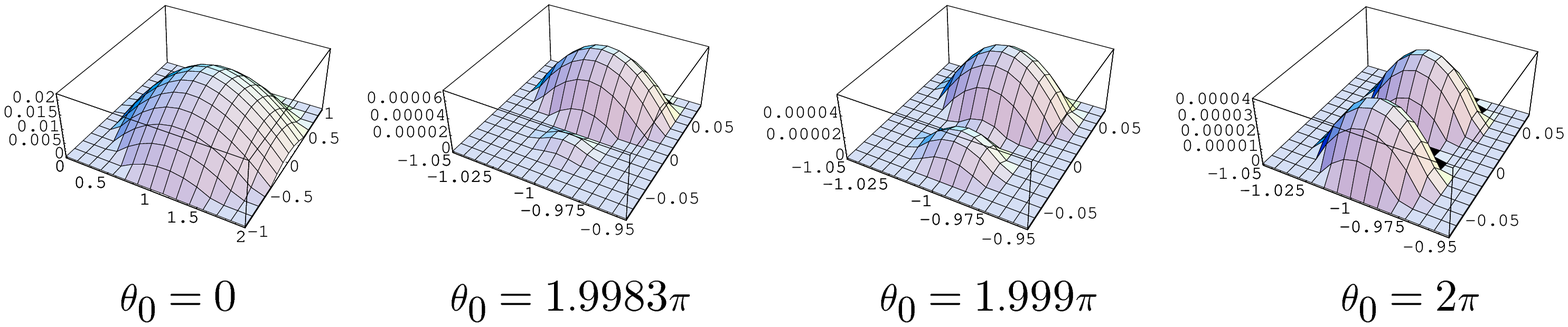}}  \vspace{-5cm}

\begin{center} Fig.8:Evolution of the area of moduli space with monopole
condensate as $\theta_0$ changes from 0 to $2 \pi$ \end{center}

It can be shown again analytically that there are two degenerate minima
at $\theta = \pi$ in the effective theory. As in the $N=1$ spurion case the
extra terms
induced by the soft breaking parameter $F_0$ in (\ref{bare}) have a $u
\rightarrow u^*$
symmetry. At $\theta_0 = \pi$ the the $Z_2$ symmetry of the $N=2$ potential
also
takes the form $u \rightarrow u^*$ and the whole potential is invariant
under this transformation. Any minima will therefore be replicated at $u^*$.

In fig. 9a,b we show the behavior of $\theta_{eff}$ as $\theta_0$ is varied.
The first order phase transition  at $\theta = \pi$ is again apparent.
Note the discontinuity in $\theta_{eff}$ as $\theta_0$ passes through
$\pi$. This discontinuity grows with the size of the soft breaking, as
shown in fig. 9b. \vspace{1cm}

$\left. \right.$  \hspace{-1cm}\ifig\prtbdiag{}
{\epsfxsize14truecm\epsfbox{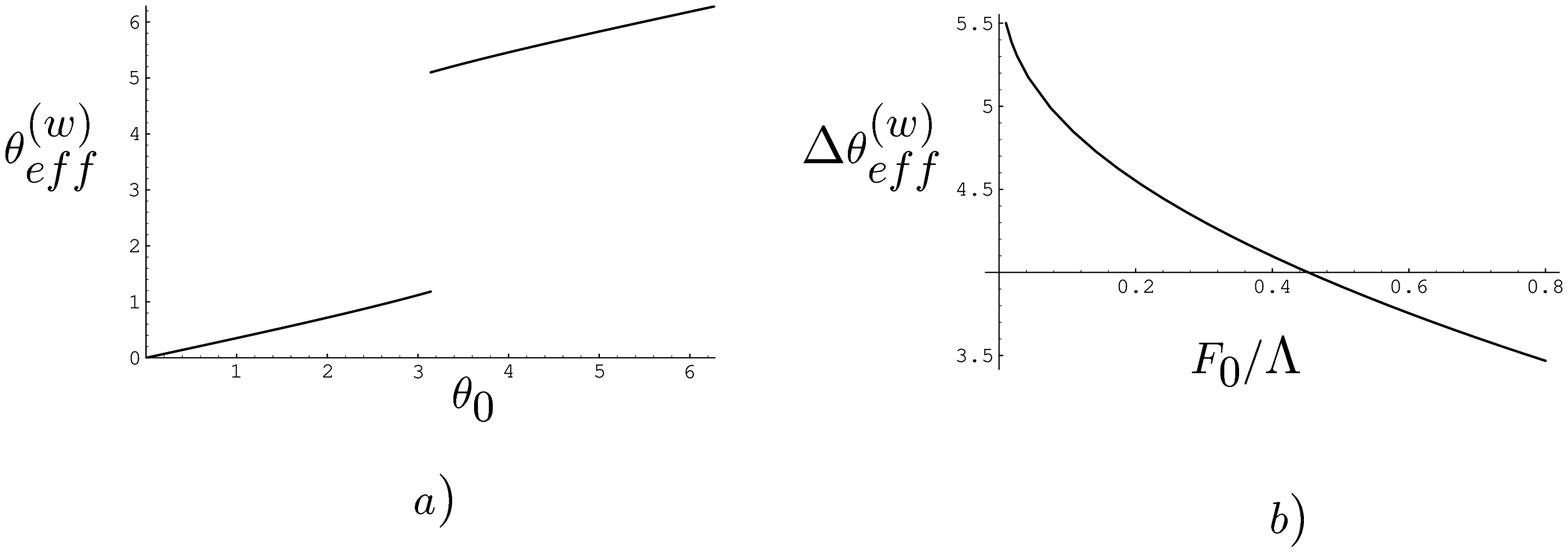}}  \vspace{-1cm}

\begin{center} Fig.9
a) $\theta^{(w)}_{eff}$ at the minimum as a 
function of  $\theta_0$ for $F_0/\Lambda
= 0.5$;
b) the discontinuity in
$\theta^{(w)}_{eff}$ at the phase transition
as a function of the soft breaking
 parameter $F_0/\Lambda$.
 \end{center}

\section{Conclusions}

In this paper we studied the effects of a non-zero
$\theta_0$ angle on softly broken $N=2$ SQCD. The
soft breakings were necessary because in their absence
the $\theta_0$ parameter is unphysical and can be rotated away
through an anomalous $U(1)_R$ symmetry.
We studied two types of controllable breakings, resulting
from $N=1$ and $N=2$ supersymmetric spurion analysis.

The general phenomena we observed is illustrated in fig. 10
below. Once the soft breakings are added,
the degeneracy of the $N=2$ moduli space is lifted
and the theory is pinned at a global minimum near {\it one} of the
singular points at $u_{\pm} = \pm \Lambda^2$. Since the
$Z_2$ symmetry is broken by the perturbation, the energy at one
of the singular points is slightly higher than at the other.
To see the origin of this
dependence on $\theta_0$, one notes that $\Lambda^2$, regarded
as a spurion field itself, has a non-trivial dependence on $\theta_0$
through the beta function (see equation (\ref{Lam2})). As $\theta_0$ is
varied from zero to $2 \pi$, there is a first order phase transition
at a critical value of $\theta_0$ in which the minimum jumps from
near one singular point to the other. In the $N=2$ spurion case,
as well as in the $N=1$ case with $f_m$ real, the
critical value is $\theta_0 = \pi$.

Since the effective theta angle is a function of
$\theta_0$ the mode of confinement changes with $\theta_0$;
it occurs by  the condensation of
dyons with charges
\beq
\label{charge}
 (q_e,q_m)~=~({\theta^{(w)}_{eff} \over 2\pi},1)~~, ~~
({\theta^{(w)}_{eff} \over 2\pi} + 1,1)~~.
\eeq
For small values of $\Delta \theta^{(w)}_{eff}$,
the discontinuity in $\theta_{eff}^{(w)}$ between the two phases,
this means that
the transition at $\theta_0 = \pi$ is between phases with condensates of
charges approximately $(0,1)$ and $(1,1)$.
For $\Delta \theta_{eff} = 0$,
the two phases are related by T-duality
(\ref{Tdual}).  However, for non-zero
$\Delta \theta_{eff}$ the phases are physically
distinct.
Using the properties of hypergeometric functions,
one can also show for both the $N=1$ and $N=2$ spurion models that
when $\theta_0 = \pi$,
\beq
\label{delta}
\theta_{eff}^{(w)} (u) +  \theta_{eff}^{(w)} (u^*) = - 2 \pi~~.
\eeq

In the
context of QCD,
't Hooft  \cite{TH} suggested that two distinct types electrically and
magnetically charged
dyons could co-exist at $\theta = \pi$ leading him to
propose that a condensate of a bound state of the two dyons would
be energetically prefered, leading to `oblique confinement'.
In the models investigated here though, for small soft breaking, the
two regions with different dyon condensates are well separated on the
$u$ plane and hence oblique confinement is not possible.
However, in \cite{N2S}, it has been argued that in the QCD limit,
$\vert F_0 \vert \rightarrow \infty$, the vacuum exhibits
simultaneous condensates of both types of excitations
(\ref{charge}).

\vskip -1cm

$\left. \right.$  \hspace{-2cm}\ifig\prtbdiag{}
{\epsfxsize8truecm\epsfbox{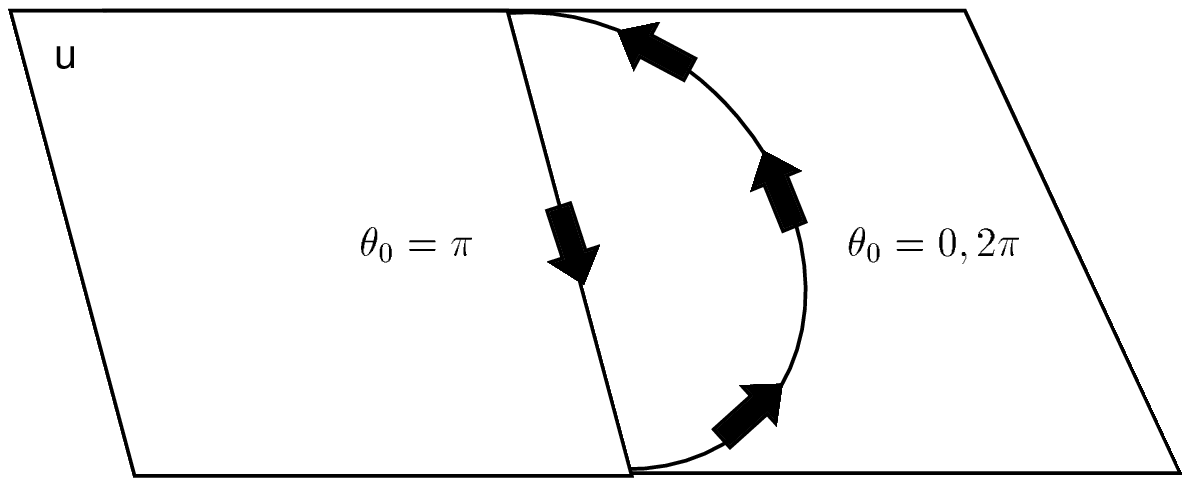}}  \vspace{-2cm} \\

\begin{center} Fig.10: Movement of minima with changing $\theta_0$
in the $N=1,2$ spurion models. \end{center}

\noindent 
The relation (\ref{delta}) then implies that when $\theta_0 = \pi$
the two types of dyons can form bound states of charge
$(0,2)$ (i.e. purely magnetically charged).
This phenomena is reminiscent of
oblique confinement.

Although we have concentrated on $N=2$ SQCD without
matter fields in the fundamental representation (i.e. $N_f = 0$),
it is also possible to address the case of two massless flavors
of quarks in the fundamental representation using the same
analysis \cite{SW2,N2S}. When $N_f = 2$
and the quarks are massless, the elliptic curve specifying the
low-energy solution is identical to that of the $N_f = 0$ case,
as long as one rescales the charges so that the $W^{\pm}$ bosons
have charge $\pm 2$ rather than $\pm 1$. Thus, one can describe
the physics of the Coulomb phase (i.e. with vanishing fundamental
squark vevs) of this model using the results
already described in this paper. The main difference is the relation
between the scale $\Lambda^2$ and $\theta_0$. By noting the
charge rescaling, or by integrating the beta function
$b_ 0 = 2 N_c - N_f$, one obtains
\beq
\Lambda^2 ~\sim~ e^{i \theta_0}~.
\eeq
This implies that as $\theta_0$ is varied from zero to $2 \pi$,
the $N_f = 2$ model undergoes {\it two} first order phase
transitions, at $\theta_0 = \pi /2, 3 \pi / 2$.

Finally, we make some speculative
remarks concerning the QCD limit,
which requires taking the soft breakings to be large compared
to $\Lambda$. In this limit the minimum of the potential must
approach $u \simeq 0$, apparently
forcing $\theta_{eff} \rightarrow \pi$
regardless of the value of $\theta_0$.
This is interesting because the effective
theta angle felt by the dyonic excitations is then independent of
the initial value of the bare theta angle.
Unfortunately, things are somewhat more complicated than
this because there may be additional corrections to the
effective theta angle which come from higher dimension
D-terms such as
$\int d^4 \theta~ F_0^{\dagger} F_0 / \Lambda^2 ~ (D W)^2$.
These are of course unsuppressed in the QCD limit, so we
do not know precisely how the low-energy theta angle behaves
as we continue to increase the soft breakings. However,
the complicated relation between $\theta_{eff}$ and $\theta_0$
exhibited in the models studied here suggests that something
similar, and more subtle than usually imagined, may
also occur in QCD.

\vskip 1in
\newpage
\centerline{\bf Acknowledgements}
\vskip 0.1in
The authors would like to thank K. Konishi for
useful discussions.
This work was supported under
DOE contract DE-AC02-ERU3075.



\vskip 0.5in
\baselineskip=1.6pt
\begin{thebibliography}{99}
%
%
\def\np#1#2#3{  {Nucl. Phys. #1} (19#3) #2}
\def\pl#1#2#3{  {Phys. Lett. #1} (19#3) #2}
\def\pr#1#2#3{   {Phys. Rev. #1} (19#3) #2}
\def\prep#1#2#3{ {Phys. Rep. #1} (19#3) #2)}
\def\prl#1#2#3{ {Phys. Rev. Lett. #1} (19#3) #2}
%
\bibitem{TH}
G.~`t Hooft, Phys. Scr.~ {\bf 24}, 841 (1981); Phys. Scr.~
 {\bf 25}, 133 (1981);
Nucl. Phys. {\bf B190}, 455 (1981).

\bibitem{CHPT}
M.~Creutz,  Phys. Rev. ~{\bf D52}, 2951 (1995) ;

N.~Evans, S. D.H.~Hsu, A. Nyffler and M.~Schwetz (in preparation).

\bibitem{SCHZ}
G.~Schierholz, Theta vacua, confinement and the continuum limit,
Talk given at LATTICE 94: 12th International Symposium on Lattice Field Theory,
Bielefeld, Germany, 27 Sep - 1 Oct 1994.
Nucl. Phys. {\bf B}, Proc. Suppl. 42, 270 (1995).

\bibitem{SW}
N.~Seiberg and E.~Witten, Nucl. Phys. {\bf B426}, 19 (1994).

\bibitem{SW2}
N.~Seiberg and E.~Witten, Nucl. Phys. {\bf B431}, 484 (1994).

\bibitem{us}
N.~ Evans, S.D.H.~ Hsu and
M. Schwetz, Phys. Lett. {\bf  B355}, 475 (1995);
N.~Evans, S. D.H.~Hsu, M.~Schwetz and  S. B.~Selipsky,
Nucl. Phys. {\bf B456}, 205 (1995).

\bibitem{N2S}
L.~\'{A}lvarez-Gaum\'{e}, J.~Distler, C.~Kounnas and M.~Mari\~{n}o, {\bf
hep-th/9604004} ;

L.~\'{A}lvarez-Gaum\'{e} and M.~Mari\~{n}o, {\bf hep-th/9606191}.

\bibitem{THM}
G.~`t Hooft, in:
 Proc.~Europ.~Phys.~Soc.~Conf.~on High Energy Physics (1975) p.1225 ;

S.~Mandelstam, Phys. Rev.~{\bf D19}, 2391 (1979).

\bibitem{WE}
E.~Witten, Phys.Lett.~{\bf B86}, 283 (1979)

\bibitem{Konishi}
M.~Di~Pierro and K.~Konishi, {\bf hep-th/9605178}.

\bibitem{Matone}
M.~Matone, Phys.Lett.~{\bf B357}, 342, (1995)

\end{thebibliography}
\end{document}